\documentclass[twocolumn,showpacs,preprintnumbers,amsmath,amssymb]{revtex4}

\usepackage{graphicx}
\usepackage{dcolumn}
\usepackage{bm}

\begin{document}

\preprint{APS/123-QED}

\title{Dark matter sterile neutrinos in stellar collapse: alteration of energy/lepton number transport
and a mechanism for supernova explosion enhancement}

\author{Jun Hidaka}
 \email{jhidaka@ucsd.edu}
\author{George M. Fuller}%
 \email{gfuller@ucsd.edu}
\affiliation{%
Department of Physics, University of California, San Diego, La Jolla, CA 92093-0319
}%



\date{\today}


\begin{abstract}
We investigate matter-enhanced Mikheyev-Smirnov-Wolfenstein (MSW) 
active-sterile neutrino conversion 
in the $\nu_e \rightleftharpoons \nu_s$ channel in the
collapse of the iron core of a pre-supernova
star. For values of sterile neutrino rest mass $m_s$ and 
vacuum mixing angle $\theta$ (specifically, $0.5\,{\rm keV}< m_s<10\,{\rm keV}$ 
and $\sin^22\theta> 5\times{10}^{-12}$) which 
include those required for viable sterile neutrino dark matter, our one-zone
in-fall phase collapse calculations show a significant reduction in 
core lepton fraction. This would result in a smaller 
homologous core and therefore a smaller initial shock energy, disfavoring
successful shock re-heating and the prospects for an explosion. However,
these calculations also suggest that the MSW resonance energy 
can exhibit a minimum located between the center and surface of the
core. In turn, this suggests a post-core-bounce mechanism to enhance neutrino transport
and neutrino luminosities at the core surface and thereby augment shock re-heating:
(1) scattering-induced or coherent MSW
$\nu_e\rightarrow\nu_s$ conversion
occurs deep in the core, at the first MSW resonance, 
where $\nu_e$ energies are large ($\sim 150$
MeV); (2) the high energy $\nu_s$ stream outward at near light speed; (3)
they deposit their energy when they encounter the second MSW resonance
$\nu_s\rightarrow\nu_e$ just below the proto-neutron star surface. 
\end{abstract}

\pacs{14.60.Pq,95.35.+d,97.60.Bw,98.80.-k}
\maketitle

\section{Introduction}

The nature of the dark matter and the explosion mechanism for core
collapse supernovae are among the most vexing outstanding problems in
modern astrophysics. In this paper we point out a curious link between
these problems. We show how sterile neutrinos with rest mass and
vacuum mixing parameters in a range which makes them 
viable cold and warm dark matter
candidates \cite{DM1,DM2,XSF,AFP,DH,AF,Kev,BiermannKusenko,Kev2,AbazajianKoushiappas} also can alter energy and lepton number transport physics
in collapsing stellar cores, possibly completely altering
the standard core collapse/explosion model. 

In broad brush, the electron fraction $Y_e$ is lowered,
lowering the pressure and, hence, the homologous
core mass. This augers against an explosion
because the initial bounce shock energy is reduced, leading to the
shock stalling closer to the neutron star where shock re-heating by neutrino energy deposition will be less effective. However, 
we also point out here that 
alterations in the rate and nature of neutrino energy transport
engendered by active-sterile neutrino flavor conversion
could go in the direction of helping the explosion
by increasing neutrino energy
luminosities at a critical epoch when the fate of the supernova shock
is determined. 

When large magnetic fields are present, active-sterile neutrino flavor transformation involving sterile neutrinos with rest mass $m_s \sim {\rm keV}$ can facilitate the generation of asymmetric neutrino emission from the neutron star surface. This, in turn, can generate large \lq\lq kicks,\rq\rq\ which may be required to explain the space motions of at least some pulsars\cite{kicks,kicks2}. Recent work suggest that these kicks can augment the hydrodynamic transport of neutrino energy to the base of the stalled supernova shock, thereby increasing shock re-heating and helping to generate a supernova explosion\cite{FK}.

The prospect of sterile neutrinos aiding the supernova
explosion is a concept that challenges the conventional wisdom
that neutrino energy diverted into sterile states is energy unavailable
to aid shock re-heating. However, it must be kept in mind that the energy
(kinetic plus optical) in the explosion, $\sim {10}^{51}\,{\rm ergs}$,
is only $\sim 10\%$ of the neutrino sea energy immediately after core bounce,
and only a miniscule $\sim 1\%$ of the $\sim {10}^{53}\,{\rm ergs}$ in the neutrino
sea a few seconds after core bounce ($t_{\rm pb} > 2\,{\rm s}$). 

Perhaps energy is not the problem; but
the {\it transport} of neutrino energy well may be the crux problem. Here we show that if neutrinos spend some of their time in sterile states, their transport mean free paths can be re-normalized upward,
thereby increasing the neutrino luminosity at the neutrino sphere.
A modest increase in neutrino luminosity during the epoch when the
stalled bounce shock is being re-heated 
($t_{\rm pb} < 1\,{\rm s}$) may be the difference between
an explosion and a dud. Though the existence
of light sterile neutrinos may be unlikely, the tremendous
leverage they would have in altering core lepton numbers and transport properties 
suggests that we set aside theoretical prejudice for the purpose of exploration.

Active-active channel neutrino flavor mixing, which we know exists, may also
affect supernova physics. Indeed, $\nu_e \rightarrow \nu_{\mu, \tau}$ flavor transformation in the in-fall phase of stellar collapse was investigated \cite{FMWS} shortly after the discovery of the Mikheyev-Smirnov-Wolfenstein (MSW) \cite{MSW,MSW_1} mechanism for in-medium enhancement of neutrino flavor conversion. However, because of the large lepton degeneracy and low entropy in the collapsing core, only un-physically large active neutrino masses can undergo significant matter-enhanced flavor transformation there. 

The mass-squared differences, and two of the four vacuum mixing
parameters, relevant for the 3-active-neutrino ($\nu_e, \nu_\mu, \nu_\tau$) case now are
known accurately from experiment \cite{nuexpt,nuexpt_1}. The mass-squared differences are:
the \lq\lq solar neutrino scale\rq\rq\ $\delta m^2 \approx 8\times{10}^{-5}\,{\rm eV}^2$; and the \lq\lq atmospheric scale\rq\rq\ $\delta m^2\approx 3\times{10}^{-3}\,{\rm eV}^2$. These are small on the scale of neutrino mass-squared difference necessary to influence core in-fall epoch physics \cite{FMWS}.

However, the anomalous result of
the LSND experiment \cite{LSND,LSND_1,LSND_2}, if interpreted as arising from vacuum neutrino
oscillations, would seem to suggest the existence of a fourth neutrino
which, on account of the $Z^0$-width limit, must have interactions
considerabley weaker than those of the weak interaction.
This is then a so-called \lq\lq sterile neutrino,\rq\rq\ perhaps an SU$(2)$
Standard Model singlet. If true, this is an astounding result which calls into
question the \lq\lq See-Saw\rq\rq\ model for how very massive
right-handed neutrino states force neutrino states closely associated
with active flavors to be light. The mini-BooNE experiment \cite{mini} is
currently probing the same parameter range as the LSND experiment (and
more). 

Regardless of the outcome of this experiment, sterile neutrinos with
rest masses, and vacuum mixing angles with active species, in the 
ranges $0.5\,{\rm keV} < m_s < 15\,{\rm keV}$ and $\sin^22\theta \ge10^{-12}$, respectively, remain a possibility. Currently the only constraints on this speculative sector of particle physics come from X-ray astronomy\cite{x-ray,x-ray_1}.  The existence of neutrinos with these mass/mixing parameters could subtly or, more likely, drastically alter supernova core and shock re-heating physics as well as the supernova neutrino signal. Lighter sterile neutrinos have been considered in other aspects of the supernova environment\cite{CFQ,1995hep.ph...11323P,1997PhRvD..56.1704N,1999PhRvC..59.2873M,Fetter:2002xx}. The central point of this paper is that to assess the effects of these sterile neutrinos on supernovae we must include them starting from the outset of core collapse, as many \lq\lq down-stream\rq\rq\ aspects of composition, equation of state, and energy/entropy transport which affect the explosion mechanism, nucleosynthesis, and expected neutrino signal could be affected.

In Section II we discuss active-sterile neutrino flavor conversion through MSW resonances in the collapsing core. We examine the prospects for energy and lepton number transport enhancement in Section III. We give a discussion and conclusions in Section IV. Appendix A describes the liquid drop model equation of state and the lepton number re-distribution schemes employed in our one-zone collapse code.

\section{Resonant Active-Sterile Neutrino Flavor Conversion in Stellar Collapse}

\subsection{The Standard In-Fall Epoch Collapse Scenario}

The weak interaction in general affects nearly every aspect of the core collapse supernova phenomenon. Neutrino emission, scattering, and absorption significantly influence the core collapse epoch and the prospects for obtaining a viable supernova explosion. It is therefore no surprise that massive sterile neutrinos could grossly alter the standard model for supernova core collapse.

Stars with initial masses $> 10\,{\rm M}_\odot$ evolve through a succession of nuclear burning phases, dumping entropy along the way through neutrino emission, eventually forming a Chandrasekhar mass ($\sim 1.4\,{\rm M}_\odot$) core supported by relativistically degenerate electrons, possessing a low entropy $s\approx 1$ (in units of Boltzmann's constant per baryon), and composed of iron-peak nuclei in nuclear statistical equilibrium (NSE). This core will go dynamically unstable and will collapse to nuclear density on a time scale $\sim 1\,{\rm s}$. 

Low entropy, electron degenerate conditions determine the course of the weakly interacting sector during the collapse. Electron capture, which is the forward reaction in
\begin{equation}
e^-+p \rightleftharpoons n+\nu_e,
\label{ecap}
\end{equation}
proceeds rapidly because the electron Fermi energy is large, $\mu_e \approx 11.1\,{\rm MeV} {\left( \rho_{10} Y_e \right)}^{1/3}$, where $Y_e$ is the number of electrons per baryon and $\rho_{10}$ is the density in units of ${10}^{10}\,{\rm g}\,{\rm cm}^{-3}$. Because the entropy is low, the proton targets for electron capture reside principally in heavy nuclei. At first the $\nu_e$'s produced by electron capture escape and carry away entropy. However, electron capture can leave daughter nuclei in excited states with excitation energies well above the temperature $T$, thereby increasing the entropy per baryon. On balance, the entropy will rise modestly, but will always remain low, $s\sim 1$, while the temperature will be in the range $T \approx 1\,{\rm MeV}$ to $2\,{\rm MeV}$ \cite{BBAL,Fuller82}. (Here we will measure entropy $s$ in units of Boltzmann's constant per baryon.) 

As the collapse proceeds and the density increases, $Y_e$ will decrease and the mean nuclear mass $A$ will rise. Coherent neutral current neutrino scattering on nuclei has a cross section that scales like $A^2$, so eventually the $\nu_e$'s produced by electron capture will be trapped - their mean free paths will drop below the core size or, equivalently, the neutrino diffusion time will exceed the collapse time scale. Neutrino coherent neutral current scattering on heavy nuclei is essentially conservative. As a result, neutrino-electron and neutrino-neutrino scattering as well as de-excitation of hot nuclei into neutrino-antineutrino pairs must intervene to cause the $\nu_e$ distribution function to evolve into a thermal, Fermi-Dirac form. 

Eventually all lepton species will come into thermal and chemical equilibrium with the nuclear component.  This is beta equilibrium. When beta equilibrium obtains, the rates of the forward and reverse processes in Eq.\ (\ref{ecap}) become equal. These rates also will be large compared to the rates of collapse, transport, and composition change. The $\nu_e$ fraction is defined to be $Y_{\nu_e}= (n_{\nu_e}-n_{\bar\nu_e})/n_b$, {\it i.e.,} the net number of electron neutrinos minus antineutrinos per baryon. The corresponding $\nu_e$ chemical potential is $\mu_{\nu_e} \approx 11.1\,{\rm MeV} {\left(2\rho_{10} Y_{\nu_e}\right)}^{1/3}$. In beta equilibrium we will have
\begin{equation}
\mu_e - \mu_{\nu_e} = \hat{\mu} +\delta m_{n p},
\label{saha}
\end{equation}
where the difference of the neutron and proton kinetic chemical potentials is $\hat{\mu} \equiv \mu_n -\mu_p$ and the neutron-proton mass difference is $\delta m_{n p} \approx 1.293\,{\rm MeV}$.

Neutrino trapping sets in for core densities $\rho_{10}\approx 10$ to $100$, and the system attains beta equilibrium shortly thereafter. In the standard model of supernova physics, core collapse is adiabatic subsequent to neutrino trapping and the net electron lepton number (lepton fraction) $Y_L =Y_{\nu_e}+Y_e$ changes little. At neutrino trapping the electron fraction might be, for example, $Y_e \approx 0.35$. However, as the collapse proceeds and the density rises, beta equilibrium shifts. By the time the core reaches nuclear density and the collapse is halted (\lq\lq core bounce\rq\rq ), the initial lepton number has been re-distributed between electrons and $\nu_e$'s, so that $Y_e \approx 0.30$ and $Y_{\nu_e} \approx 0.05$. 

The shock wave generated at core bounce will have an initial energy that is comparable to the in-fall kinetic energy and, therefore, scales as $\sim Y_e^{10/3}$ \cite{Fuller82}. Note that any process which violates lepton number conservation and turns $\nu_e$'s into neutrinos of other flavors, including sterile flavors, will open phase space for electron capture. Electron capture prior to trapping and the establishment of beta equilibrium will lower $Y_e$, thereby lowering the initial shock energy. The additional electron capture on nuclei will increase the entropy to some extent. This will be off-set by post-trapping entropy loss if $\nu_e \rightarrow \nu_s$ results in the production of sterile neutrinos which leave the core.  
 
\subsection{The Neutrino Forward Scattering Potential and the MSW Resonance Condition}

An electron neutrino propagating through the stellar medium will experience a potential stemming from forward scattering on particles that carry weak charge (electrons, neutrons, protons, neutrinos). There will be no contribution to this potential from forward $\nu_e - \nu_{\mu,\tau}$ scattering as long as the net muon and tau lepton numbers residing in the supernova neutrino seas are zero. Absent large scale neutrino flavor conversion, this is a good approximation throughout the epochs of interest in the supernova environment. Electron capture produces $\nu_e$'s, as outlined above, but $\bar\nu_e$'s and neutrinos and antineutrinos with mu and tau flavors can appear only through thermal pair production processes and these are suppressed by the large electron Fermi energies and low entropies characteristic of the in-fall epoch. Pair production of these species will be efficient after core bounce when the entropy is larger.

With these provisos, we can express the $\nu_e$ forward scattering potential ({\it cf.,} Ref.\ \cite{AFP}) as 
\begin{eqnarray}
V & = & \sqrt{2} G_{\rm F} {\left( n_e - {{1}\over{2}} n_n \right)}+2\sqrt{2} G_{\rm F} \left( n_{\nu_e}-n_{\bar\nu_e} \right) \nonumber \\
 &  & +  \sqrt{2} G_{\rm F} \left( n_{\nu_{\mu}}-n_{\bar\nu_{\mu}} \right) 
      +  \sqrt{2} G_{\rm F} \left( n_{\nu_{\tau}}-n_{\bar\nu_{\tau}} \right) \\
& = & {{3 \sqrt{2}}\over{2}}\, G_{\rm F}\, n_b {\left( Y_e + {{4}\over{3}} Y_{\nu_e} -{{1}\over{3}}\right)},
\label{pot}
\end{eqnarray}
where $G_{\rm F}$ is the Fermi constant and $n_b = \rho N_A$ is the baryon density, with $\rho$ the density in ${\rm g}\,{\rm cm}^{-3}$ and where $N_A$ is Avogadro's number and in the last equality we have set the net muon and tau neutrino number densities to zero. Here $n_{\nu_e}$ and 
$n_{\bar\nu_e}$ are effective $\nu_e$ and $\bar\nu_e$ number densities, respectively, so that
$n_{\nu_e}-n_{\bar\nu_e} = n_b Y_{\nu_e}$. Likewise, the net electron number density $n_e$ and total (free plus nucleus) proton number density $n_p$ are related through overall charge neutrality,
$n_p = n_e = n_{e^-}-n_{e^+}=n_b Y_e$. The total neutron density is then $n_n=n_b-n_p$.

We will posit that in vacuum the unitary relation between the propagating neutrino energy (\lq\lq mass\rq\rq ) eigenstates, $\vert \nu_1\rangle$ and $\vert \nu_2\rangle$, and the weak interaction (flavor) eigenstates is
\begin{eqnarray}
\vert \nu_e\rangle & = & \cos\theta \vert \nu_1\rangle + \sin\theta \vert\nu_2\rangle \
\\
\vert \nu_s\rangle & = & -\sin\theta \vert \nu_1\rangle + \cos\theta \vert \nu_2\rangle,
\label{unitary}
\end{eqnarray}
where $\vert\nu_s\rangle$ is the sterile neutrino state and $\theta$ is an {\it effective} $2\times 2$ vacuum mixing angle. There would likely be mixing between the sterile neutrino and all three active neutrinos in vacuum, but in this work we will treat $\nu_e - \nu_s$ transformations exclusively. We argue that this at least will serve to flesh out the main features of active-sterile conversion in collapse. This is probably a good approximation during in-fall because the small measured values of neutrino mass-squared differences imply that active-active neutrino flavor conversion will be suppressed at densities characteristic of core collapse. Later, however, after core bounce, our treatment will be only approximate, as neutrino flavor mixing in this regime will be $3\times 3$ in nature. We can avoid the full $4\times 4$ neutrino flavor mixing problem 
because of the near maximal mixing between $\nu_\mu$ and $\nu_\tau$ and the near identical interactions of these species and their antiparticles in the supernova environment \cite{BF,CFQ}, at least until active-sterile and active-active neutrino flavor conversion has produced unequal and substantial net mu and tau lepton numbers. (See the discussion of the similar physics in the early universe in Ref.\ \cite{ABFW}.)

The effective {\it in-medium} $2\times 2$ mixing angle $\theta_{\rm M}$ depends on the local neutrino forward scattering potential $V$ and can be found from
\begin{equation}
\sin^22\theta_{\rm M} = {{\Delta^2 \sin^22\theta}\over{{\left( \Delta \cos2\theta -V \right)}^2+\Delta^2 \sin^22\theta  }},
\label{thetaM}
\end{equation}
where, in terms of vacuum mass eigenvalue-squared difference $\delta m^2 \equiv m_2^2-m^2_1$ and neutrino energy $E_\nu$, we define $\Delta \equiv \delta m^2/2E_\nu$. The effective in-medium mixing angle is maximal ($\theta_{\rm M}=\pi/4$) at an MSW resonance. A neutrino with energy $E_\nu$ and experiencing a forward scattering potential $V$ will be MSW-resonant when
\begin{equation}
\Delta \cos2\theta = V.
\label{MSWcond}
\end{equation}
The high densities encountered in stellar collapse dictate that at least one of the neutrinos must have a rest mass in the $\sim {\rm keV}$ range in order for there to be an MSW resonance for a typical (mean) neutrino energy \cite{FMWS}. As a result, we can safely set $\delta m^2 \approx m_2^2 \equiv m_s^2$, where $m_s$ is the rest mass associated with the sterile neutrino state. Furthermore, x-ray and closure constraints imply that the effective active-sterile mixing angle is small for this sterile neutrino mass range, so that we can set $\cos2\theta \approx 1$. With these approximations, the MSW resonance energy is
\begin{eqnarray}
E_{\rm res} & \approx & {{m_s^2}\over{ 3 \sqrt{2}\, G_{\rm F} \rho N_A {\left( Y_e + {{4}\over{3}} Y_{\nu_e} -{{1}\over{3}}\right)} }}\
\\
& \approx & {{ 4.37\,{\rm MeV} { \left( {{m_s}/{ {\rm keV}}} \right) }^2  }\over{ \rho_{12} {\left( Y_e + {{4}\over{3}} Y_{\nu_e} -{{1}\over{3}}\right)}   }}.
\label{MSWsweep}
\end{eqnarray}
We follow convention and give the matter density $\rho$ scaled by an appropriate value, so that 
$\rho_{\rm n} \equiv \rho/{10}^{\rm n}\,{\rm g}\,{\rm cm}^{-3}$. 

\subsection{Coherence and Adiabaticity}

Conversion of active neutrinos into sterile neutrinos can be facilitated either by active neutrino scattering-induced de-coherence or by coherent neutrino propagation through an MSW resonance. The former process will dominate the production of sterile neutrinos at high density, where the scattering rate is large. For the neutrino mass/mixing parameters of interest here, the latter process will dominate at lower density. Let us consider this process first.

The fate of a neutrino propagating coherently through an MSW resonance depends on two factors. 
The first factor is obviously coherence itself. By \lq\lq coherence\rq\rq\ we mean coherence on the scale of the MSW resonance width. Therefore, coherent neutrino evolution will correspond to large values ($> 1$) of the ratio of the neutrino's mean free path in medium to the MSW resonance width. 

The second issue in the neutrino's fate has to do with the degree to which neutrino flavor evolution is adiabatic. \lq\lq Adiabaticity\rq\rq\ is gauged by the ratio of the resonance width and the neutrino flavor oscillation length at resonance. Adiabatic evolution means that the neutrino remains in its particular superposition of initial instantaneous mass eigenstate(s) as it propagates through resonance.

{\it Complete} conversion of the neutrino's flavor will result if both of these ratios are large compared to unity and if the far asymptotic initial state of the neutrino corresponds closely to both an instantaneous mass eigenstate and a flavor eigenstate. For the densities and neutrino mass/mixing parameters of interest in stellar collapse, efficient coherent production of sterile neutrinos at an MSW resonance is always the result when the neutrino mean free path is large compared to the resonance width and the oscillation length at resonance is small compared to this width.

The width of the resonance is $\delta t = {\cal{H}} \tan2\theta$, where the effective weak charge scale height is ${\cal{H}} \equiv {\vert d{\ln V}/dt\vert}^{-1}$, and where $t$ can be either a time or length parameter along the neutrino's world line. The adiabaticity parameter $\gamma$ for a neutrino with energy $E_{\rm res}$ is proportional to the ratio of $\delta t$ and the neutrino oscillation length at resonance
$L^{\rm osc}_{\rm res} =4\pi E_{\rm res}/(\delta m^2 \sin2\theta)$, and is defined to be
\begin{equation}
\gamma \equiv 2\pi {{\delta t}\over{L^{\rm osc}_{\rm res}}} = {{\delta m^2 {\cal{H}}  }\over{ 2 E_{\rm res} }} \cdot {{\sin^22\theta }\over{ \cos2\theta }}.
\label{adiabat}
\end{equation}

Prior to neutrino trapping, $\nu_e$'s produced by electron capture will propagate at near light speed along radomly-directed trajectories. After trapping, the $\nu_e$'s will have nearly isotropic propagation directions. In either regime, however, sterile neutrinos $\nu_s$ will freely stream at near light speed. The $\nu_s$'s will be produced with randomly distributed propagation directions, reflecting the distribution of $\nu_e$ propagation directions.

Here we will consider sterile neutrinos which are directed radially outward. For this case, ${\cal{H}}$ computed along the radial direction is the relevant  potential scale height. More tangentially-directed sterile neutrinos will experience an effective potential scale height which is larger than that in the radial direction. As a result, our calculations with radially-directed neutrinos will give {\it lower} limits on adiabaticity and, hence, will give under-estimates of neutrino flavor conversion efficiency. 

\begin{figure}[htbp]
\includegraphics[width=3.4in]{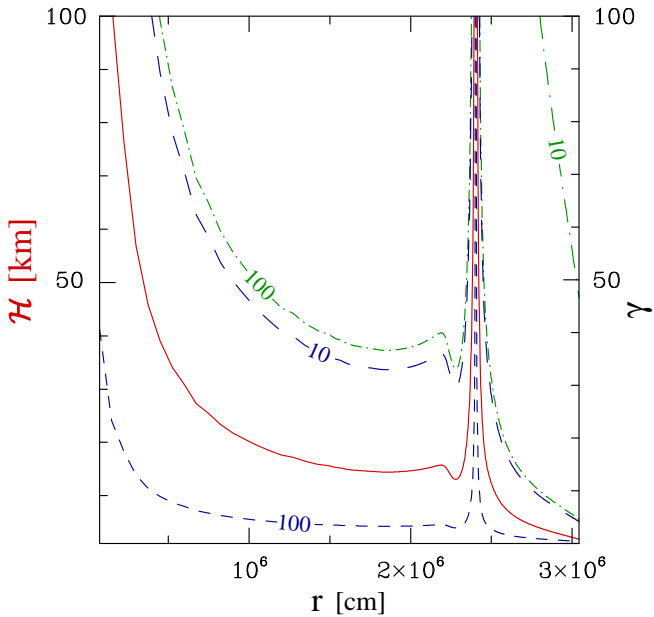}
\caption{\label{fig:density_scale_height} The solid curve shows the effective potential scale height $\mathcal{H}$ (left vertical axis) in the radial direction as a function of radius $r$.   Also shown is the adiabaticity parameter $\gamma$ (right vertical axis) as a function of $r$ for $\nu_e\rightleftharpoons\nu_s$ with $\sin^22\theta={10}^{-9}$ and with sterile neutrino rest mass $m_s = 3\,{\rm keV}$ and energy $E_\nu = 10\,{\rm MeV}$ (long-dashed curve) and $E_\nu = 100\,{\rm MeV}$ (short-dashed curve), and for sterile neutrinos with rest mass $m_s = 10\,{\rm keV}$ and energy $E_\nu = 10\,{\rm MeV}$ (long dash-dot curve) and $E_\nu = 100\,{\rm MeV}$ (short dash-dot curve).} 
\end{figure}

Figure~\ref{fig:density_scale_height} shows the potential scale height ${\cal{H}}$ for radially-directed neutrino trajectories as a function of radius. Also shown in this figure is the adiabaticity parameter $\gamma$ relevant for sterile neutrinos with $E_\nu = 10\,{\rm MeV}$ and $E_\nu=100\,{\rm MeV}$ and with $m_s =3\,{\rm keV}$ and $m_s = 10\,{\rm keV}$, all for an assumed vacuum mixing angle in the $\nu_e\rightleftharpoons\nu_s$ channel satisfying $\sin^22\theta={10}^{-9}$.
In general, lower energy neutrinos have smaller oscillation lengths at resonance and, consequently, larger adiabaticity parameters. This is evident in the figure. The \lq\lq spike\rq\rq\ in potential scale height and adiabaticity parameter evident at radius $r\approx 24\,{\rm km}$ in this figure, corresponds to a maximum in the potential $V$ to be discussed below.

After neutrino trapping, when $\nu_e$'s have relatively short mean free paths, the mean density experienced by a $\nu_e$ will rise with time as the collapse proceeds. In this case, we can follow
Ref.~\cite{FMWS} and view the MSW resonant flavor conversion in the mean rest frame of the $\nu_e$. In our one-zone collapse calculations we take the collapse rate to be, (see Appendix A)
\begin{equation}
\label{collaprate}
{{d \ln\rho}\over{dt}}=\left(100\,{\rm s}^{-1}\right) \sqrt{\rho_{10}},
\end{equation}
and with this choice the adiabaticity parameter appropriate
for $\nu_e \rightarrow \nu_s$ is
\begin{equation}
\label{collapad}
\gamma = {{\delta m^2}\over{200 E_\nu \rho_{10}^{1/2}}}.
\end{equation}

At the very high densities encountered in the core near and after bounce, scattering-induced de-coherence can result in significant production of sterile neutrinos (see for example the discussion in Ref.\ \cite{AFP}). The sterile neutrino emissivity in this case will be proportional to the product of the active neutrino scattering rate, a factor which takes account of the quantum Zeno effect, and $\sin^22\theta_M$. This latter term will again make the sterile neutrino production rate largest at MSW resonances, where $\sin^22\theta_M =1$. The quantum Zeno factor is inversely proportional to the active neutrino scattering rate when this rate is large. This behavior tends to accentuate the enhancement of sterile neutrino production at MSW resonances.

\subsection{Evolution of the Potential}

We calculate the coupled histories of the neutrino forward scattering potential $V$, the electron fraction $Y_e$, and the $\nu_e$ fraction $Y_{\nu_e}$ with a one-zone collapse/equation-of-state code described in Appendix~A. This code includes a mean nucleus approximation and a self consistent calculation of entropy/temperature and electron capture rates on free nucleons and heavy nuclei. The density-time/radius histories of zones located at different radius will be self similar, so long as the collapse is homologous (in-fall velocity proportional to radius). With homology, all Lagrangian mass zones in the core will experience similar time-density histories, albeit with different beginning and ending points. In this case, the density of any zone at the instant of core bounce (which we take to be when the central density reaches $\rho_{12}=3\times{10}^2$) corresponds to the central density at an earlier time. Of course, electron capture and the associated fall in pressure will cause the actual collapse to deviate from homology. Nevertheless, our one-zone code serves to illustrate the qualitative details important in understanding active-sterile neutrino flavor transformation and its effects on the equation of state in the core. 

\begin{figure}[htbp]
\includegraphics[width=3.4in]{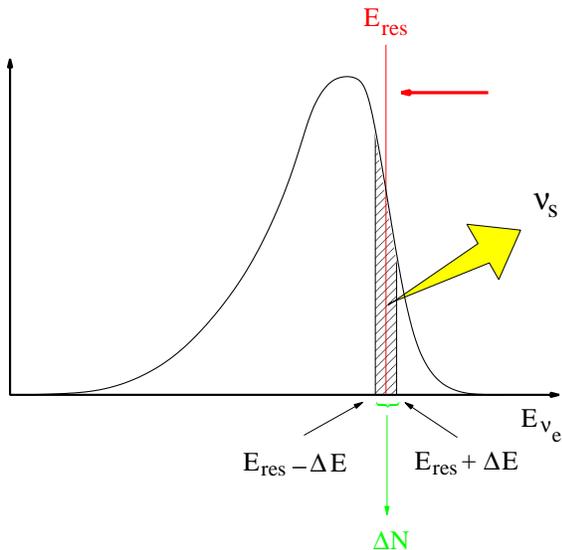}
\caption{\label{fig 2}The resonance energy $E_{\rm res}$ sweeps from higher toward lower energy through the $\nu_e$ distribution function (shown as a function of energy $E_{\nu_e}$), at any instant converting the $\nu_e$'s contained in energy interval $E_{\rm res}-\Delta E$ to $E_{\rm res}+\Delta E$ (where $\Delta E = E_{\rm res}\,\tan2\theta$) to sterile neutrinos $\nu_s$. The number of $\nu_e$'s converted at this instant is $\Delta N$.} 
\end{figure}

The rise in density with time during collapse dictates the evolution of $V$ and $E_{\rm res}$. The potential $V$ rises and $E_{\rm res}$ falls with increasing density ({\it i.e.,} Eqs.~\ref{MSWsweep} and \ref{pot}). The MSW resonance in this case sweeps from higher toward lower energy through the $\nu_e$ distribution which, after neutrino trapping and the attainment of beta equilibrium, we can regard as Fermi-Dirac in character, $f_{\nu_e}\left( E_\nu\right) = {\left[ T^3 F_2\left(\eta\right)\right]}^{-1} E_\nu^2/\left(e^{E_\nu/T-\eta}+1\right)$, where $T$ is the temperature, $\eta$ is the electron neutrino degeneracy parameter $\mu_{\nu_e}/T$, and the relativistic Fermi integrals of order $k$ are defined by $F_k\left(\eta\right) \equiv \int_0^\infty{  {{x^2\,dx}\over{e^{x-\eta}+1}}}$. This is depicted in Fig.~\ref{fig 2}. 
If neutrino evolution is adiabatic, at any instant all the $\nu_e$'s within the MSW width $\Delta E = E_{\rm res}\,\tan2\theta$ will be converted to sterile neutrinos.

\begin{figure}[htbp]
\includegraphics[width=3.4in]{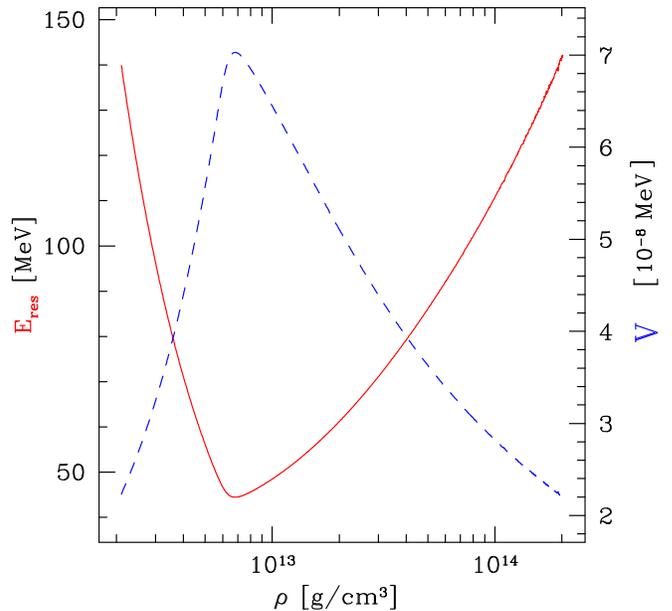}
\caption{\label{fig 3}One-zone calculations of $\nu_e\rightleftharpoons\nu_s$ resonance energy $E_{\rm res}$ in MeV (left vertical axis, solid line) and $\nu_e$ forward scattering potential $V$ in units of ${10}^{-8}\,{\rm MeV}$ (right vertical axis, dashed line) are shown as a function of density (in ${\rm g}\,{\rm cm}^{-3}$). This calculation employs $m_s= 3\,{\rm kev}$ and $\sin^22\theta={10}^{-9}$}
\end{figure}

The one-zone calculation of $\nu_e$ forward scattering potential and resonance energy as a function of density for $m_s= 3\,{\rm kev}$ and $\sin^22\theta={10}^{-9}$ and for the constant collapse rate discussed above is shown in Figure~\ref{fig 3}. 
The striking feature in this plot is the peak in potential and corresponding minimum in resonance energy.

The explanation for this peak/minimum is straightforward. The trend of potential $V$ with increasing density changes dramatically when the MSW resonance energy $E_{\rm res}$ sweeps low enough to reach the $\nu_e$ Fermi energy, $\mu_{\nu_e}$. Given the degenerate conditions, there are not many $\nu_e$'s with energies above the Fermi energy $\mu_{\nu_e}$. Early in the collapse, when $E_{\rm res} > \mu_{\nu_e}$, not many sterile neutrinos are produced and there is only a modest decrease in $Y_e$. However, once $E_{\rm res} \le \mu_{\nu_e}$, substantial numbers of $\nu_e$'s can be converted to sterile $\nu_s$'s and this can have a large effect on electron capture.. 

Neutrino flavor conversion $\nu_e \rightarrow \nu_s$ opens holes in the $\nu_e$ distribution which allows for more electron capture and a concomitant decrease in electron lepton number per baryon $Y_l = Y_e+Y_{\nu_e}$. The portion of the change in lepton number stemming from the decrease in the number of electrons, $\Delta Y_e$, and the portion stemming from the decrease in the number of $\nu_e$'s, $\Delta Y_{\nu_e}$, is calculated self-consistently with the equation of state as described in Appendix~A. Most of the lepton number decrease will be in $Y_e$, but some of it will be in $Y_{\nu_e}$. In any case, the potential $V$ is proportional to the factor $Y_e+{{4}\over{3}} Y_{\nu_e}-{{1}\over{3}}$, and this factor will decrease when there is substantial electron capture, {\it e.g.,} when $E_{\rm res} \le \mu_{\nu_e}$. The net result is that $V$ will {\it decrease} with increasing density beyond this point.

\begin{figure}[htbp]
\includegraphics[width=3.4in]{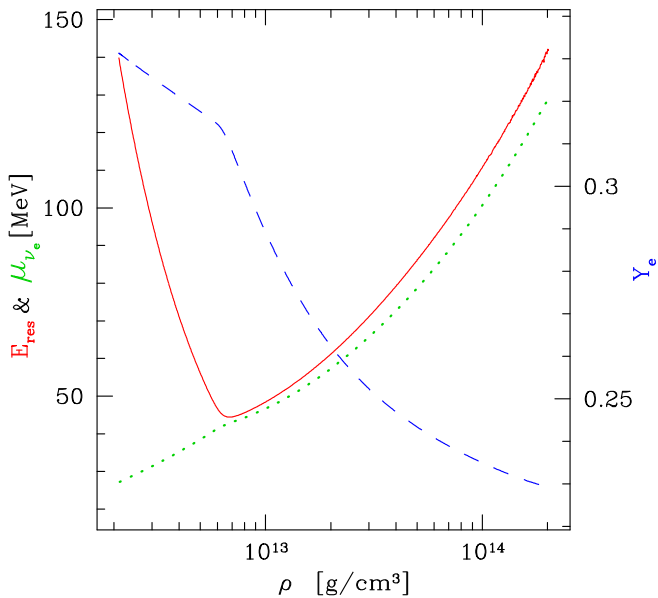}
\caption{\label{fig 4} Same one-zone calculation as in Figure~\ref{fig 3}: resonance energy $E_{\rm res}$ (solid line), electron fraction $Y_e$ (dashed line), and $\nu_e$ Fermi energy $\mu_{\nu_e}$ (dotted line) are shown as functions of density.} 
\end{figure}

In fact, there is a rough feed-back that develops once the resonance energy reaches the $\nu_e$ Fermi energy. Increasing density would tend to decrease $E_{\rm res}$, but this is counteracted by a decrease in the factor $Y_e+{{4}\over{3}} Y_{\nu_e}-{{1}\over{3}}$. The result is that the resonance energy tracks the $\nu_e$ Fermi energy $\mu_{\nu_e} \approx 11.1\,{\rm MeV} {\left(2\rho_{10} Y_{\nu_e}\right)}^{1/3}$, increasing with increasing density and, more or less, staying just above $\mu_{\nu_e}$. This is evident in Figure~\ref{fig 4}. This figure shows $E_{\rm res}$ as a function of density for a one-zone calculation with the same parameters as in Figure~\ref{fig 3}, but now also shows the corresponding values of the $\nu_e$ Fermi energy $\mu_{\nu_e}$ and the electron fraction $Y_e$.

The calculation of electron capture and, {\it e.g.,} $Y_e$ in the one-zone calculations depends on an accurate estimate of neutrino flavor conversion in the $\nu_e\rightarrow\nu_s$ channel. In turn, as outlined above, this depends on the degree to which neutrino flavor evolution is coherent and adiabatic. There are three time scales in our one-zone calculation that give a rough idea of when these conditions obtain: $T_{\rm coll}$,
$T_{\rm osc}$, and $\delta t_{\rm res}$. $T_{\rm coll}$ is the mean-free-time between collisions, $T_{\rm osc}\equiv L_{\rm osc}^{\rm res}$ is the oscillation time at resonance, and $\delta
t_{\rm res}$ is the resonance width appropriate for core in-fall as defined above. We then have

\begin{equation}
T_{\rm coll}=\frac{1}{\sigma_n n_{\rm target} v_{\rm rel}},
\label{eq:t_coll} 
\end{equation}
\begin{equation}
T_{\rm osc}\approx \frac{4 \pi E_{\nu_e}}{m_s^2 \sin 2\theta},
\label{eq:t_osc} 
\end{equation}
\begin{equation}
\delta t_{\rm res}=\frac{1}{100\sqrt{\rho_{10}^{}}}
\frac{\delta\rho_{10}^{}}{\rho_{10}^{}}
=\frac{1}{100\sqrt{\rho_{10}^{}}}\tan2\theta.
\label{eq:dt_res} 
\end{equation}
Here $\sigma_n$ is the cross section for a $\nu_e$ of energy $E_{\nu_e}$ to scatter on a neutron. We take this cross section to be
\begin{equation}
\sigma_n=\frac{1}{4}\sigma_0 \left(\frac{E_{\nu_e}}{m_e c^2}\right)^2,
\label{eq:sigma_n} 
\end{equation}
where
\begin{equation}
\sigma_0=\frac{4 G_F^2 m_e^2 \hbar^2 }{\pi c^2}\approx 1.76\times 10^{-44}\,{\rm cm}^2.
\label{eq:sigma_0} 
\end{equation}
In fact, the neutrino-nucleus coherent scattering cross section (crudely, larger than $\sigma_n$ by nuclear mass-squared, $A^2$) sets the mean free path, so our calculations here over-estimate $T_{\rm coll}$.
In Eq.\ (\ref{eq:t_coll}), $n_{\rm target}$ is the number density of nucleon/nucleus targets and
$v_{\rm rel} \approx c$ is the relative velocity between $\nu_e$ and
target nucleons/nuclei.  Coherent, adiabatic flavor conversion $\nu_e\rightarrow\nu_s$ would require that the condition $T_{osc} \ll \delta t_{res} \ll T_{coll}$ be satisfied.

\begin{figure}[htbp]
\includegraphics[width=3.1in]{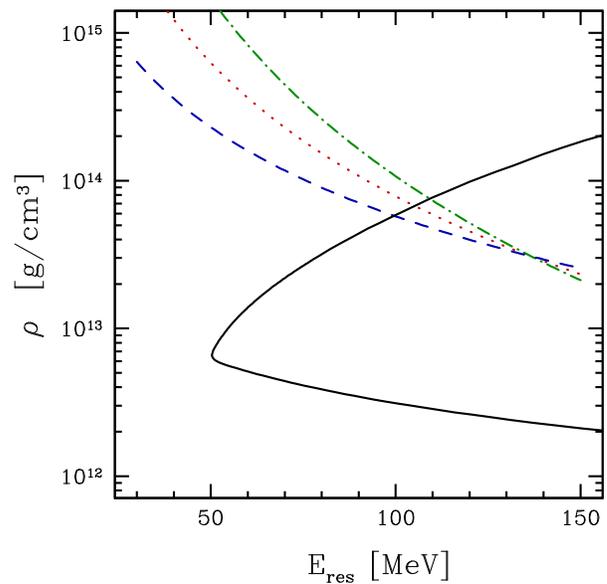}
\caption{\label{fig 5} The one-zone calculation evolutionary track (solid line) with $m_s=3\,{\rm keV}$ and $\sin^22\theta={10}^{-9}$ is shown in the resonance energy $E_{\rm res}$, density $\rho$ (${\rm g}\,{\rm cm}^{-3}$) plane. The conditions $T_{\rm osc}/\delta t_{\rm res}=1$, $T_{\rm osc}/T_{\rm coll}=1$ , and $\delta t_{\rm res}/T_{\rm coll}=1$ obtain along the dashed, dotted, and dot-dashed contours, respectively. Values of these ratios are $< 1$ everywhere below these contours. } 
\end{figure}

The validity of this condition for the particular track of density, composition, and resonance energy in a one-zone calculation with $m_s=3\,{\rm keV}$ and $\sin^22\theta={10}^{-9}$ is depicted in Figure~\ref{fig 5}. In this figure contours where $T_{\rm osc}/T_{\rm coll}=1$, $\delta t_{\rm res}/T_{\rm coll}=1$, and $T_{\rm coll}/\delta t_{\rm res}=1$ are shown as functions of resonance energy $E_{\rm res}$ and density $\rho$ (in ${\rm g}\,{\rm cm}^{-3}$). These ratios are $< 1$ below these contours. We conclude that coherence and adiabaticity is likely for all but the highest densities, near or beyond core bounce. Though these estimates are crude and leave out details of, {\it e.g.,} detailed active neutrino cross section physics, we can conclude conservatively that neutrino flavor evolution can be regarded as coherent and adiabatic for sterile neutrino parameters of interest for dark matter, at least through the epoch where the MSW resonance energy begins to track the $\nu_e$ Fermi energy. This epoch begins at the \lq\lq knee\rq\rq\ evident in the one-zone evolutionary track in Figure~\ref{fig 5}. This is the key issue in $Y_e$ evolution in the collapsing core.

\subsection{Electron Fraction Reduction}

\begin{figure}[htbp]
\includegraphics[width=3.4in]{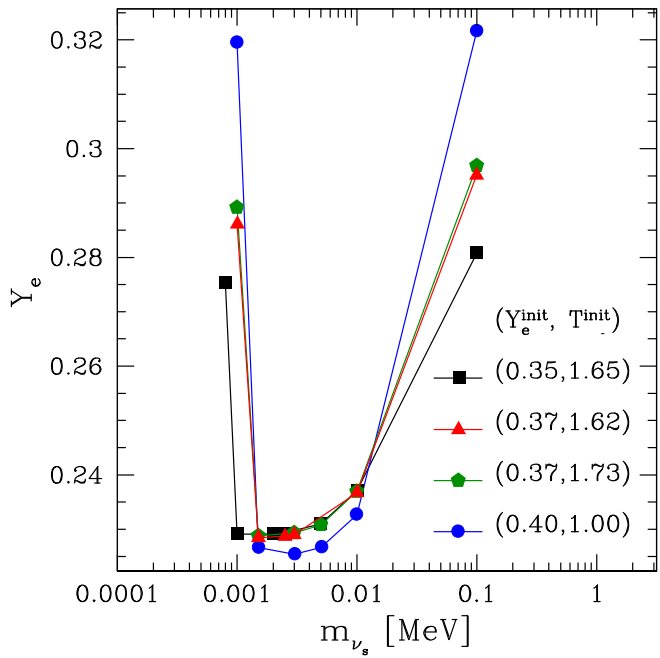}
\caption{\label{fig 6}Final electron fraction $Y_e$ is shown as a function of sterile neutrino rest mass $m_{s}$ as calculated in a one-zone code with $\sin^22\theta={10}^{-9}$ and initial values of electron fraction $Y_e^{\rm init}$ and temperature $T^{\rm init}$ (in MeV) as shown.} 
\end{figure}

The most striking change in core composition in the presence of $\nu_e\rightarrow\nu_s$ is the reduction in electron fraction $Y_e$. In Figure~\ref{fig 6} we show results of our one-zone calculations with $\sin^22\theta={10}^{-9}$. This figure gives the final (at core bounce) value of $Y_e$ as a function of sterile neutrino rest mass (in MeV) for several different values of initial electron fraction $Y_e^{\rm init}$ and temperature $T^{\rm init}$ (in MeV) at the onset of collapse.
In broad brush, our calculations show that for sterile neutrino rest mass in the range $1\,{\rm keV} \le m_s < 10\,{\rm keV}$ and $\sin^22\theta > {10}^{-12}$, a significant subset of the range of parameters of interest for dark matter, there is a substantial reduction in electron fraction, down to final values $Y_e \approx 0.22$ from an initial value in the range $Y_e^{\rm init} =0.35$ to $0.4$. This is in stark contrast to a one-zone calculation without $\nu_e\rightarrow\nu_s$, which would give a final electron fraction value $Y_e \approx 0.33$ for the same range of initial electron fraction. 

Since, as outlined above, the homologous core at bounce forms the \lq\lq piston\rq\rq\ for the shock, the initial shock energy scales like $\sim Y_e^{10/3}$. As a result, the $\sim 30\%$ reduction in final electron fraction relative to the case without sterile neutrinos translates into more than a factor of two reduction in initial shock energy. 

And that is just the beginning of the problem. With a smaller homologous core there will be more material for the shock to traverse in the outer core. This is because the outer core is the remainder of the initial iron core which is not in the homologous, inner core.  The entropy jump across the shock will be $\Delta s \sim 10$ units of Boltzmann's constant per baryon, implying a shift in composition in Nuclear Statistical Equilibrium from heavy nuclei in front of the shock to free nucleons and alpha particles behind it. This process is sometimes termed nuclear photo-dissociation. 

Since nucleons are bound in nuclei by $\sim 8\,{\rm MeV}$ on average, the shock will lose ${10}^{51}\,{\rm ergs}$ per $0.1\,{\rm M}_\odot$ of outer core material traversed. Furthermore, our one-zone calculations show little electron capture-induced increase in core entropy during collapse relative to the no-sterile-neutrinos standard model. In part, this is because most of the extra electron capture stemming from holes in the $\nu_e$ distribution opened by $\nu_e\rightarrow\nu_s$ occurs when the medium is already in beta equilibrium. Also, sterile neutrinos escaping directly from the core remove entropy. On balance, our one-zone calculations show only small net changes in entropy per baryon.

The large sterile neutrino-induced decrease in initial shock energy and increased nuclear photo-dissociation burden may mean that shock stagnation occurs at a smaller radius than in standard core collapse models. In turn, with the shock stalled at lower radius, there would be less optical depth to neutrino absorption below the shock and, consequently, one would expect neutrino re-heating to be less efficacious. In short, the reduction in $Y_e$ engendered by $\nu_e\rightarrow\nu_s$ augers against a explosion.

At this point one might be tempted to conclude that since (some) core collapse events lead to manifest explosions, the sterile neutrino mass/mixing parameters for which there is a significant $Y_e$ reduction are ruled out. This conclusion is ill advised for two reasons: (1) we do not know the detailed physics of shock propagation and electron capture/neutrino transport in the post-bounce regime; and, (2) sterile neutrinos can affect the {\it rate} at which lepton number and energy/entropy is transported to the region behind the shock. 

For example, based on a proper account of electron capture on very heavy neutron-rich nuclei in the post-bounce epoch, recently it has been shown that the stall point of the shock may be relatively {\it insensitive} to the value of $Y_e$ at bounce \cite{Hix}. This is despite the arguments to the contrary given above! The only robust conclusion that we can draw is that a large sterile neutrino-induced $Y_e$ reduction will alter significantly the standard model of core collapse. These alterations would be on a scale far in excess of those stemming from nuclear physics and hydrodynamics/transport issues that modelers currently deal with.

The second issue, sterile neutrino alteration of energy/entropy/lepton number transport, is the subject of the next section.

\section{Enhancement of Energy and Lepton Number Transport}

The mechanism behind the explosion of core collapse supernovae is not well understood. The energy for the explosion comes ultimately from gravitational energy, but the way this energy is transported seems to be crucial \cite{2004ApJS..150..263L,2003ApJ...592..434T,2005ApJ...620..840L,2005AAS...207.1701M,2006ApJ...642..401B,2004rpao.conf..224L,2005ApJ...626..317W,2002ApJ...574L..65F,2005PhRvD..72d3007C,2001ApJ...560..326B,2006astro.ph..7281S}. Contemporary two and three-dimensional models show explosions \cite{2006ApJ...642..401B,2002ApJ...574L..65F,1995ApJ...450..830B,2006A&A...447.1049B,2006A&A...453..661K}, while spherically symmetric, one-dimensional models, which boast the most sophisticated transport, equation of state, and nuclear physics do not seem to yield convincing explosions. Perhaps this reflects the actual situation in nature. However, many uncertainties in the fundamental physics input remain in these models and it is an open question whether vigorous explosions will occur in one dimension \cite{OpenIssuesBook}.

Simulations suggest that a modest increase in neutrino energy luminosity from the proto-neutron star surface (neutrino sphere) might increase neutrino energy deposition rates behind the shock enough to cause a viable explosion in what would otherwise have been a dud. This shows that neutrino energy transport in the proto-neutron star core may be the crux issue in the one-dimensional core collapse explosion mechanism problem. 

\begin{figure}[htbp]
\includegraphics[width=3.4in]{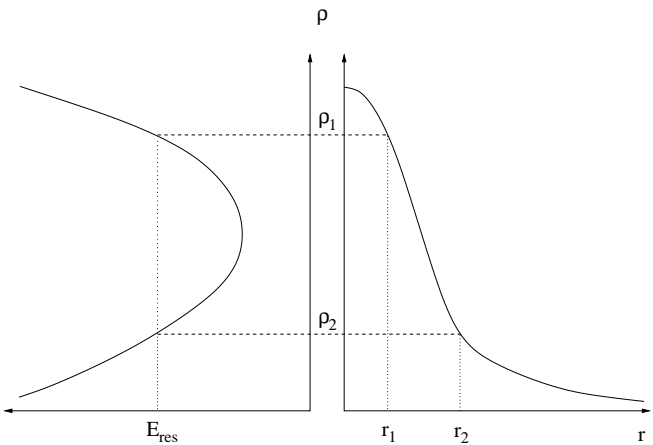}
\caption{\label{fig 7} Resonance energy (left) and radius parameter $r$ (right) in the in-falling, pre-bounce core are shown back-to-back as functions of density $\rho$ (vertical axes). An example given resonance energy $E_{\rm res}$ corresponds to two locations, $r_1$ and $r_2$, and two corresponding densities, $\rho_1$ and $\rho_2$.} 
\end{figure}

The peak in the potential $V$ and associated dip in the resonance energy $E_{\rm res}$ found in our calculations suggests a process whereby the $\nu_e$ neutrino luminosity might be increased. With the local minimum in $E_{\rm res}$ near the middle of the core, a neutrino of a given energy could encounter two MSW resonances: one deep in the core and one further out, closer to the neutrino sphere.  Figure~\ref{fig 7} shows the resonance energy and radius parameter in the in-falling, pre-bounce core as functions of density.

Imagine that there is a source of sterile neutrinos from flavor conversion occurring deep in the core, where $\nu_e$ energies are $\sim \mu_{\nu_e}$ and so are large, perhaps $\sim 100\,{\rm MeV}$.  This sterile neutrino emissivity could arise from medium-enhanced coherent flavor conversion $\nu_e\rightarrow\nu_s$ at the inner MSW resonance or, more likely, scattering-dominated de-cohrerent flavor conversion in any channel $\nu_\alpha\rightarrow\nu_s$ ($\alpha=e,\mu,\tau$) or medium-enhanced, de-coherent $\nu_e\rightarrow\nu_s$ occurring at the inner MSW resonance. 

\begin{figure}[htbp]
\includegraphics[width=2.5in]{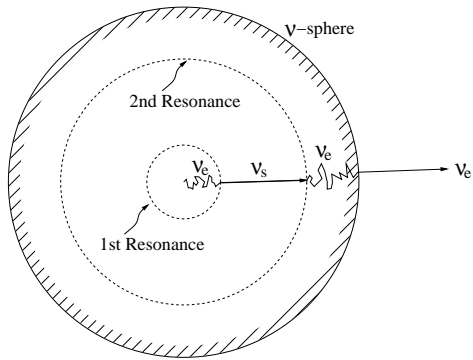}
\caption{\label{fig 8} High energy $\nu_e$'s could be converted to sterile neutrinos deep in the core and then re-generated as $\nu_e$ further out, nearer the neutrino sphere (edge of core).} 
\end{figure}

These high energy sterile neutrinos would move out at near light speed and encounter the second, outer MSW resonance. There they could be coherently re-converted $\nu_s\rightarrow\nu_e$ to active $\nu_e$'s. The result would be an upward re-normalization of the net electron lepton number and energy transport rates. In other words, flavor mixing would allow $\nu$ neutrinos to spend part of the time in sterile states that have higher effective diffusion rates.  This is depicted in Figure~\ref{fig 8}.

For this mechanism to aid the shock re-heating/explosion process , the rough arrangement of an inner source of sterile neutrinos at high energy and an outer MSW resonance would have to survive shock propagation through the core. This is because the epoch at which shock re-heating occurs is between a few hundred milliseconds and $1\,{\rm s}$ after core bounce. The shock is generated at the edge of the inner core and will alter the entropy and, hence, the composition in the outer core.

However, some gross features of our shock energy enhancement scheme are easy to discern. If, for example, the feedback mechanism relating $\nu_{e}\rightarrow\nu_{s}$ flavor conversion and potential $V$ continues to hold through the epoch when the shock propagates through the outer core, then, as in the in-fall epoch, the resonance energy $E_{\rm res}$ will track the $\nu_{e}$ chemical potential $\mu_{\nu_e}$ and will remain slightly above it, $E_{\rm res}\gtrsim\mu_{\nu_e}$. If in addition, $\mu_{\nu_e}$ simply tracks the density so that near the center of the core $\mu_{\nu_e}\sim 150\,{\rm MeV}$, then nearly all the sterile neutrinos created by scattering-induced de-coherence deep in the core will be re-converted to active $\nu_e$ neutrinos further out. Taking the high energy $\nu_s$ emissivity from Ref.~\cite{AFP} and assuming complete conversion of this energy into $\nu_e$'s which thermalize quickly nearer the core's edge, we estimate that $\lesssim 10^{52}\,{\rm ergs}$ of additional neutrino energy could be radiated from the neutrino sphere during the $\sim 1\,{\rm s}$ duration of shock re-heating. This is, however, a very crude estimate. We will estimate the effects of shock propagation on the neutrino flavor transformation potential in a subsequent work.

\section{Discussion and Conclusions}

We have demonstrated that the existence of sterile neutrinos with rest masses in the range of interest for dark matter could have significant implications for the in-fall, collapse epoch of core collapse supernovae. In particular, sterile neutrinos with rest masses in the range  $1\,{\rm keV} \le m_s < 10\,{\rm keV}$, and possessing effective vacuum mixing angles characteristic of the channel $\nu_e\rightleftharpoons\nu_s$ which satisfy $\sin^22\theta > {10}^{-12}$, could result in enhanced electron capture during in-fall with a concomitant decrease in electron fraction $Y_e$. As discussed in Section II~E, the decrease in $Y_e$ will translate into a decrease in relativistic electron degeneracy pressure in the core. In turn, the consequence of this will be a smaller homologous core and, therefore, a lower initial shock energy and an increased photo-disintegration burden on the shock. Although these alterations certainly auger against obtaining a viable shock and explosion in the framework of current models, it must be kept in mind that there remain significant uncertainties in fundamental nuclear electron capture, equation of state, and transport physics in the post core bounce regime.

\begin{figure}[htbp]
\includegraphics[width=3.1in]{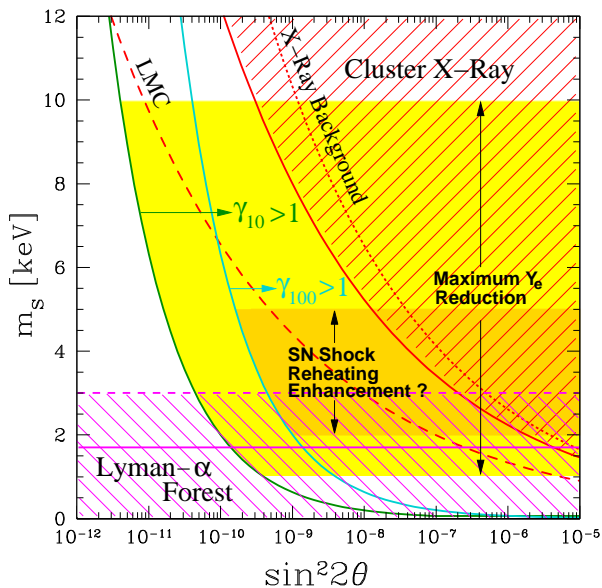}
\caption{\label{fig 9}{\small Lightly shaded and heavily shaded regions show the ranges of sterile neutrino rest mass $m_s$ (in keV) and effective $\nu_e\rightleftharpoons\nu_s$ vacuum mixing $\sin^22\theta$ that give large $Y_e$ reduction and shock re-heating/explosion enhancement, respectively.
The line labeled by $\gamma_{100}>1$ shows the contour of mixing parameters along which the adiabaticity parameter $\gamma$ at the outer MSW resonance is unity or greater for neutrinos with energy $E_\nu = 100\,{\rm MeV}$. The corresponding line $\gamma_{10}>1$ shows where $\gamma$ is unity or greater for neutrinos with energy $E_\nu=10\,{\rm MeV}$. The heavy solid line and the cross-hatched region to the right show the range of parameters ruled out by X-ray observations of the Virgo cluster \cite{AFP,x-ray}. The dotted line and the region to its right is ruled out by observations of the diffuse X-ray background \cite{2006MNRAS.370..213B}. The dashed line (labeled LMC) is the claimed X-ray limit from the Large Magellanic Cloud \cite{Boyarsky}. The dashed and solid horizontal lines are the Lyman-$\alpha$ forest-derived lower limits on $m_s$ from Ref.~\cite{Viel} and Ref.~\cite{Abazajian06}, respectively.}} 
\end{figure}

However, we could conservatively conclude that the existence of sterile neutrinos with these properties would result in a supernova paradigm radically different from current models. Additionally, we have outlined a means whereby active-sterile-active neutrino flavor transformation (specifically, $\nu_e\rightarrow\nu_s\rightarrow\nu_e$) could alter electron lepton number and energy transport rates. Active neutrinos which spend part of their time propagating as sterile neutrinos will have larger transport mean free paths on average. In short, much of supernova physics \lq\lq down stream\rq\rq\ of core bounce could be quite different from standard models.

For example, a higher $\nu_e$ luminosity as a result of $\nu_e\rightarrow\nu_s\rightarrow\nu_e$ would result in a number of changes in the near hydrostatic envelope above the neutrino sphere and below the shock. This would mean a much reduced neutron excess in this region, since the neutron-to-proton ratio is determined locally by a competition between $\nu_e+n\rightleftharpoons p+e^-$ and $\bar\nu_e+p\rightleftharpoons n+e^+$. This would have implications for alpha-rich freeze-out, neutrino-heated nucleosynthesis coming from this region. In fact, this effect could be beneficial, increasing $Y_e$ and lowering the yield of neutron-rich, neutron number $N=50$ nuclei. However, the initial neutrino spectra at the neutrino sphere could be altered by the the enhanced deposition of $\nu_e$'s just below the proto-neutron star surface. This could partially counteract the neutron excess-lowering effect in the envelope by reducing the average $\nu_e$ energy relative to that for $\bar\nu_e$. 

In the end, our results call into question the old idea that sterile neutrinos remove energy from the core and, therefore, cannot exist, else there would be no supernova explosions. It must be kept in mind that the supernova explosion energy (kinetic plus optical) is $\sim {10}^{51}\,{\rm erg}$, which is only $\sim 1\%$ of the total amount of energy radiated away over some $10\,{\rm s}$ as neutrinos of all kinds. It may be possible to \lq\lq throw away\rq\rq\ much of this energy and still get an explosion. The timing, rate, and location of energy deposition are the most important determinants of explosion physics and we have demonstrated that some or all of these can be affected by sterile neutrinos with masses/mixings in the range of interest for dark matter.

In Figure~\ref{fig 9} we summarize the the parameter space of sterile neutrino rest masses and effective vacuum mixing angles (in the $\nu_e\rightleftharpoons\nu_s$ channel) that gives rise to supernova effects. This figure also shows possible constraints stemming from cosmological sterile neutrino radiative decay ($\nu_s \rightarrow \nu_\alpha + \gamma$). To obtain significant $\nu_e\rightarrow\nu_s\rightarrow\nu_e$ enhancement of $\nu_e$ luminosity at the neutron star surface (neutrino sphere), high energy $\nu_s$'s coming from deep in the core must encounter the outer MSW resonance and be converted with high efficiency to $\nu_e$'s. In the figure, the line labeled by $\gamma_{100}>1$ shows the contour of mixing parameters along which the adiabaticity parameter $\gamma$ at the outer MSW resonance is unity for neutrinos with energy $E_\nu = 100\,{\rm MeV}$. Likewise, this figure also shows the corresponding contour, $\gamma_{10}>1$, for adiabaticity for neutrinos with energy $E_\nu=10\,{\rm MeV}$. The lightly shaded region in this figure, which shows the parameter range for large reduction in $Y_e$, is bounded on the low mixing angle side by the contour $\gamma_{10}>1$. This is because augmentation of electron capture requires only that $\nu_e$ neutrinos with energies $E_\nu \sim 10\,{\rm MeV}$ be efficiently converted  to $\nu_s$'s.

The existence of relatively small scale structure in the universe, {\it i.e.,} the Lyman alpha forest, sets an upper limit on neutrino suppression of the matter power spectrum and the dark matter particle collisionless damping scale. In turn, this sets a lower limit on the rest mass of sterile neutrinos if they are to be the bulk of the dark matter. This limit is in dispute. In Figure~\ref{fig 9} we show both the Abazajian limit \cite{Abazajian06} and a more conservative limit \cite{Viel:2005qj}. Likewise, the lack of an observed X-ray line feature in observations of the Large Magellanic Cloud (LMC) has been argued to set an upper limit on sterile neutrino rest mass \cite{Boyarsky}. 
We also show current diffuse X-ray background constraints \cite{2006MNRAS.370..213B}. Perhaps the most stringent constraints come from the lack of observation of a sterile neutrino X-ray decay line from the Virgo cluster \cite{AFP,x-ray}. This is the heavy solid line and the cross-hatched region to its right. 

With all constraints applied, there remains a significant region of unconstrained $m_s\--\sin^22\theta$ parameter space in which there could be large supernova in-fall epoch and shock re-heating effects. Neutrinos with these parameters could also be the dark matter, at least if there is a significant primordial lepton number \cite{XSF,AFP,AF,DM2} or with particular couplings \cite{DM2}. These neutrinos fall in the range of parameters that, after core bounce, could engineer large pulsar kicks \cite{kicks2,kicks}  and modify hydrodynamic neutrino energy transport to the base of the shock \cite{FK}.

The ultimate conclusion that can be drawn from our work is that stellar collapse is remarkably sensitive to new lepton number-violating physics in the neutrino sector. The reason for this is clear: the energy in core collapse supernovae is gravitational. This gravitational binding energy is stored in degenerate seas of leptons ($\nu_e$'s and electrons). Any lepton number violating processes in operation can tap into this energy and thereby alter supernova core composition and dynamics.
Sterile neutrinos with rest masses $\sim {\rm keV}$ which mix with active neutrinos at the level of one part in $\sim {10}^{11}$ in vacuum are likely to be undetectable in conventional terrestrial experiments but, remarkably, could completely change the current model for core collapse supernovae.

\begin{acknowledgments}
This work was supported in part by NSF grant PHY-04-00359 at UCSD and
the TSI collaboration's DOE SciDAC grant at UCSD. We thank 
K. Abazajian, P. Amanik, A. Kusenko, A. Mezzacappa, M. Patel, and J. R. Wilson for valuable discussions.
\end{acknowledgments}

\appendix

\section{Numerical Calculations}
To gauge the effects of altered electron capture physics in the in-fall epoch of stellar collapse, we employ a modified version of the one-zone code described in Ref.~\cite{Fuller82}. For the neutron kinetic chemical potential (in MeV) we use
\begin{equation}
\mu_n=-16+125(0.5-Y_\epsilon)-125(0.5-Y_\epsilon)^2
-\frac{W_{\rm surf}}{A^{1/3}}\frac{3-7Y_\epsilon}{2(1-Y_\epsilon)},
\end{equation}
while for the difference of the neutron and proton kinetic chemical potentials, $\hat\mu=\mu_n-\mu_p$, (again, in MeV) we employ
\begin{equation}
\hat\mu=250(0.5-Y_\epsilon)-W_{\rm surf}A^{-1/3}\left(\frac{1}{Y_\epsilon}
+\frac{2}{Y_\epsilon}\frac{1-2Y_\epsilon}{1-Y_\epsilon}\right).
\end{equation}
In these expressions, $Y_\epsilon\equiv Y_e/(1-X_n)$, where $X_n$ is the free neutron mass fraction. Following Ref.~\cite{BBAL}, the nuclear surface energy $W_{\rm surf}$ and the nuclear Coulomb energy $W_{\rm coulomb}$ are taken to be:  
\begin{equation}
W_{\rm surf}=(290\,{\rm MeV}) Y_\epsilon^2(1-Y_\epsilon)^2;
\end{equation}
\begin{equation}
W_{\rm coulomb}=(0.75\,{\rm MeV}) Y_{\epsilon}^2(1-0.234\rho_{12}^{1/2}+0.00194\rho_{12}).
\end{equation}
The mean nuclear mass is $A$. This is determined by a minimization of nuclear energy which gives the condition $W_{\rm surf}A^{2/3}=2W_{\rm coulomb}A^{5/3}$, from which we obtain
\begin{equation}
A=194(1-Y_\epsilon)^2(1-0.234\rho_{12}^{1/2}+0.00194\rho_{12})^{-1}.
\end{equation}

With these expressions it is evident that $\mu_n$ and $\hat\mu$ are function s of $Y_e$, $X_n$, and $\rho$
\begin{equation}
\hat\mu=\hat\mu(Y_e, X_n, \rho),
\label{eq:mu_hat}
\end{equation}
\begin{equation}
\mu_n=\mu_n(Y_e, X_n, \rho).
\end{equation}
Likewise, as discussed in section II, the electron and $\nu_e$ chemical potentials are functions of $\rho$, temperature $T$ (or in energy units $k_{B}T$, where $k_B\simeq 0.08617\,\rm{MeV/T_{9}}$ is Boltzmann's constant), and $Y_e$ or $Y_{\nu_e}$, 
\begin{equation}
\mu_{\nu_e}=\mu_{\nu_e}(Y_{\nu_e}, k_B T, \rho),
\end{equation}
\begin{equation}
\mu_e=\mu_e(Y_e, k_B T, \rho ).
\end{equation}

The chemical potentials of the leptons and the free nucleons are related through the beta equilibrium condition, Eq.\ (\ref{saha}). We assume that beta equilibrium is attained on a time scale short compared to the dynamical time implied in Eq.\ (\ref{collaprate}). The beta equilibrium condition is then a constraint among $Y_e$, $Y_{\nu_e}$, $k_{B}T$, and $X_n$, which we could describe as
\begin{equation}
f_{\beta\textrm{-eqil}}(X_n,k_B T,Y_e,Y_{\nu_e},\rho)=0.
\label{eq:constraint_beta}
\end{equation}
Note that we take the time evolution of $\rho$ explicitly from Eq.\ (\ref{collaprate}).

Another constraint on the system can be obtained by assuming Nuclear Statistical Equilibrium (NSE), in which case the Saha equation gives 
\begin{equation}
X_n\simeq79\frac{(k_B T)^{3/2}}{\rho_{10}}\exp\left(\frac{\mu_n}{k_B T}\right).
\end{equation}
This is a good approximation whenever nucleons are non-degenerate, {\it i.e.}, well below nuclear saturation density at $\rho_{\rm 12}\simeq 300$. The assumption of NSE provides a constraint among $Y_{e}$, $k_{B}T$, and $X_n$, and could be written, 
\begin{equation}
f_{\rm NSE}(X_n,k_B T,Y_e,\rho)=0.
\label{eq:constraint_MB}
\end{equation}

Since particle number in both election capture and active-sterile flavor transformation $\nu_e\rightarrow\nu_s$ is preserved, we have an additional constraint on the number of spin-$1/2$ leptons, 
\begin{equation}
Y_{e}+Y_{\nu_e}+Y_{\nu_s}=Y_L^0=\rm{const},
\label{eq:lepton_conserv1}
\end{equation}
where $Y_L^0$ is the initial (at neutrino trapping) lepton number per baryon. Equivalently, we can consider the time derivative of this constraint equation, 
\begin{equation}
\dot Y_{e}+\dot Y_{\nu_e}+\dot Y_{\nu_s}=0.
\label{eq:lepton_conserv2}
\end{equation}
The sterile neutrinos are not trapped, of course, so $\dot Y_{\nu_s}$ could be thought of as a sterile neutrino number emissivity per baryon.

The entropy per baryon in our calculation is regarded to be a funcion of $Y_e$, $Y_{\nu_e}$, $X_n$, $k_{B}T$, and $\rho$, namely $S(\rho,Y_e,Y_{\nu_e},X_n,k_BT)$. Therefore, the entropy evolves in time according to 
\begin{eqnarray}
\lefteqn{S(\rho^{(\textrm{n+1})},Y_e^{(\textrm{n+1})},Y_{\nu_e}^{(\textrm{n+1})},X_n^{(\textrm{n+1})},k_BT^{(\textrm{n+1})})}\nonumber\\ 
&=S(\rho^{(\textrm{n})},Y_e^{(\textrm{n})},Y_{\nu_e}^{(\textrm{n})},X_n^{(\textrm{n})},k_BT^{(\textrm{n})})+\dot S \Delta t,
\label{eq:diff_eq_S}
\end{eqnarray}
where $\Delta t$ is a time increment related by Eq.\ (\ref{collaprate}) to the increment in density. We could write the total differential of entropy in a time/density step as 
\begin{eqnarray}
dS&=&\frac{1}{k_BT}\Big[dQ_{\nu_e\rightarrow\nu_s}\nonumber\\
  &&-\Big(\mu_e dY_e+(\mu_n+m_nc^2)dY_n\nonumber\\
  &&+(\mu_p+m_pc^2)dY_p+\mu_{\nu_e}dY_{\nu_e}\Big)\Big]\\
  &=&\frac{1}{k_BT}[dQ_{\nu_e\rightarrow\nu_s}\nonumber\\
  &&  -(\mu_e-\mu_{\nu_e}-\hat\mu+\delta m_{np})dY_e\nonumber\\
  &&+\mu_{\nu_e}dY_{\nu_s}]\\
  &=&\frac{1}{k_BT}[dQ_{\nu_e\rightarrow\nu_s}+\mu_{\nu_e}dY_{\nu_s}]
\end{eqnarray}
where $dQ_{\nu_e\rightarrow\nu_s}$ and $dY_{\nu_s}$ are the heat loss
per baryon and stelile neutrino number generation, respectively, stemming from the flavor conversion $\nu_e\rightarrow\nu_s$ occuring during the time step as the resonance energy $E_{\rm res}$ (Eq.\ (\ref{MSWsweep})) sweeps through a portion of the $\nu_e$ distribution. Note that the $\beta$-equilibrium condition (imposed to simplify the second-to-last equality above) gives a simple relation between the time rates of change of the entropy and heat loss per baryon and the sterile neutrino production rate per baryon, 
\begin{equation}
\dot S=\frac{1}{k_BT}[\dot Q_{\nu_e\rightarrow\nu_s}+\mu_{\nu_e}\dot
Y_{\nu_s}].
\label{eq:dot_S} 
\end{equation}
At an MSW resonance with physical width $\delta t_{\rm res}$ and total $\nu_e$ neutrino energy contained in this width $\Delta E_{\nu_e}(Y_{\nu_e},T)$, we have 
\begin{equation}
\dot Q_{\nu_e\rightarrow\nu_s}=-\frac{\Delta E_{\nu_e}(Y_{\nu_e},T)}{\delta
t_{\rm res}(\rho)}=\dot Q_{\nu_e\rightarrow\nu_s}(Y_{\nu_e},T,\rho),
\end{equation}
\begin{equation}
\dot Y_{\nu_s}=\frac{\Delta N_{\nu_e}(Y_{\nu_e},T)}{\delta t_{\rm res}(\rho)}
=\dot Y_{\nu_s}(Y_{\nu_e},T,\rho),
\label{eq:dotYnus} 
\end{equation}
where $\Delta N_{\nu_e}(Y_e,Y_{\nu_e},T)$ is the number of $\nu_e$'s within the resonance width. With these constraints, definitions, and simplifications, Eq.\ (\ref{eq:diff_eq_S}) becomes
\begin{eqnarray}
\lefteqn{S(\rho^{(\textrm{n+1})},Y_e^{(\textrm{n+1})},Y_{\nu_e}^{(\textrm{n+1})},X_n^{(\textrm{n+1})},k_BT^{(\textrm{n+1})})}&&\nonumber\\ 
&=&S(\rho^{(\textrm{n})},Y_e^{(\textrm{n})},Y_{\nu_e}^{(\textrm{n})},X_n^{(\textrm{n})},k_BT^{(\textrm{n})})\nonumber\\ 
&&+\dot S(Y_{\nu_e}^{(\textrm{n})},k_BT^{(\textrm{n})},\rho^{(\textrm{n})}) \Delta t.
\label{eq:S_evolution}
\end{eqnarray}
Combining the constraints in Eq.\ (\ref{eq:lepton_conserv1}) and Eq.\ (\ref{eq:dotYnus}) we obtain
\begin{equation}
Y_e^{(\textrm{n+1})}+Y_{\nu_e}^{(\textrm{n+1})}=Y_e^{(\textrm{n})}+Y_{\nu_e}^{(\textrm{n})}-\dot Y_{\nu_s}(Y_{\nu_e}^{(\textrm{n})},k_BT^{(\textrm{n})},\rho^{(\textrm{n})})\Delta t.
\label{eq:Y_evolution} 
\end{equation}
The other two constraints, Eqs.~\ref{eq:constraint_beta} and 
\ref{eq:constraint_MB}, can be expressed as
\begin{equation}
f_{\beta\textrm{-eqil}}(X_n^{(\textrm{n+1})},k_B^{(\textrm{n+1})} k_BT^{(\textrm{n+1})},Y_e^{(\textrm{n+1})},Y_{\nu_e}^{(\textrm{n+1})},\rho^{(\textrm{n+1})})=0,
\label{eq:constraint_beta1}
\end{equation}
and
\begin{equation}
f_{\rm NSE}(X_n^{(\textrm{n+1})},k_B T^{(\textrm{n+1})},Y_e^{(\textrm{n+1})},\rho^{(\textrm{n+1})})=0.
\label{eq:constraint_MB1}
\end{equation}
Eqs.\ (\ref{eq:S_evolution}), (\ref{eq:Y_evolution}),
(\ref{eq:constraint_beta1}), and (\ref{eq:constraint_MB1}) form a set
of non-linear equations with respect to variables
$Y_e^{(\textrm{n+1})}$, $Y_{\nu_e}^{(\textrm{n+1})}$,
$X_n^{(\textrm{n+1})}$, and $k_B T^{(\textrm{n+1})}$ and can be solved
numerically, {\it e.g.}, by a Newton-Raphson method.


\newpage 

\bibliography{ref_sterile_by_George}

\end{document}